\documentclass[preprintnumbers,amsmath,amssymbm,prd]{revtex4}
\usepackage{epsfig}
\usepackage{graphicx}
\usepackage{amssymb}

\begin{document}

\title{Lower bound on the compactness of isotropic ultra-compact objects}
\author{Shahar Hod}
\affiliation{The Ruppin Academic Center, Emeq Hefer 40250, Israel}
\affiliation{ } \affiliation{The Hadassah Institute, Jerusalem
91010, Israel}
\date{\today}

\begin{abstract}

\ \ \ Horizonless spacetimes describing spatially regular
ultra-compact objects which, like black-hole spacetimes, possess
closed null circular geodesics (light rings) have recently attracted
much attention from physicists and mathematicians. In the present
paper we raise the following physically intriguing question: How
compact is an ultra-compact object? Using analytical techniques, we
prove that ultra-compact isotropic matter configurations with light
rings are characterized by the dimensionless lower bound
$\text{max}_r\{2m(r)/r\}>7/12$ on their global compactness
parameter.
\end{abstract}
\bigskip
\maketitle

\section{Introduction}

Null circular geodesics on which massless particles (gravitons and
photons) can orbit a central astrophysical object are a generic
feature of highly compact matter configurations
\cite{Bar,Chan,ShTe}. In particular, the compact theorem presented
in \cite{Hodlb} has revealed the fact that spherically symmetric
black-hole spacetimes with a non-positive energy-momentum trace
($T\leq0$ \cite{Bond1,Notetr}) must posses at least one closed null
circular geodesic.

It is well known, however, that {\it horizonless} matter
configurations whose characteristic global compactness parameter
$\text{max}_r\{2m(r)/r\}$ \cite{Noteunits,Notemr} is less than that
of classical black-hole spacetimes \cite{Notecbh}, may also possess
null circular geodesics (see
\cite{Carm,Hodt0,Kei,Mag,Hodmz1,CBH,Hodmz2,Zde1} and references
therein). As possible spatially regular exotic alternatives to the
canonical black-hole spacetimes, the physical and mathematical
properties of these horizonless curved spacetimes with light rings
have attracted much attention in recent years
\cite{Kei,Hodt0,Mag,CBH,Carm,Hodmz1,Hodmz2,Zde1}.

Spatially regular matter configurations which, like black-hole
spacetimes, possess null circular geodesics are usually known in the
physics literature by the exotic name {\it ultra-compact objects}
\cite{Carm,Hodt0,Kei,Mag,Hodmz1,CBH,Hodmz2,Zde1}. In the present
paper we would like to raise the following
physically interesting question: How compact is an ultra-compact
object which possesses light rings?

In particular, one naturally wonders whether it is possible, within
the framework of general relativity, to give a more precise {\it
quantitative} meaning to the important physical term
`ultra-compact object'? In this context, it is worth mentioning that the
theorem presented in \cite{Hodt0} has revealed the fact that
spatially regular matter configurations with a non-negative
energy-momentum trace ($T\geq0$) are necessarily ultra-compact. That
is, the corresponding spherically symmetric horizonless curved
spacetimes must possess (at least) one light ring \cite{Noteift1}.
Moreover, it has been proved in \cite{Hodt0} that the characteristic
global compactness parameter of these horizonless matter
configurations is bounded from below by the simple dimensionless
relation
\begin{equation}\label{Eq1}
\text{max}_r\Big\{{{2m(r)}\over{r}}\Big\}> {2\over3}\ \ \ \
\text{for}\ \ \ \ T\geq0\  ,
\end{equation}
where $m(r)$ is the gravitational mass of the matter fields
contained within a sphere of areal radius $r$. To the best of our
knowledge, no analogous lower bound on the dimensionless global
compactness parameter $\text{max}_r\{{{2m(r)}/{r}}\}$ of
ultra-compact objects with light rings and a {\it negative}
energy-momentum trace \cite{Bond1} has thus far been presented in
the physics literature.

The main goal of the present paper is to study analytically the
physical and mathematical properties of spherically symmetric ultra-compact objects whose
horizonless spacetimes possess closed light rings (null circular
geodesics). In particular, below we shall explicitly prove that the
global compactness parameter of spatially regular ultra-compact
isotropic matter configurations with a negative energy-momentum
trace is bounded from below by the dimensionless relation
$\text{max}_r\{{{2m(r)}/{r}}\}>7/12$.

\section{Description of the system}

We study the physical properties of horizonless spatially regular
isotropic matter configurations which possess closed light rings
(null circular geodesics). These ultra-compact objects are described
by the spherically symmetric line element \cite{Chan,Notesc}
\begin{equation}\label{Eq2}
ds^2=-e^{-2\delta}(1-C)dt^2 +(1-C)^{-1}dr^2+r^2(d\theta^2
+\sin^2\theta d\phi^2)\  ,
\end{equation}
where, for spatially regular asymptotically flat spacetimes, the
radially dependent metric functions $\{C,\delta\}$ are characterized
by the small-$r$ \cite{Hodt1}
\begin{equation}\label{Eq3}
C(r\to 0)=O(r^2)\ \ \ \ {\text{and}}\ \ \ \ \delta(0)<\infty\
\end{equation}
and large-$r$ \cite{Hodt1,May}
\begin{equation}\label{Eq4}
C(r\to\infty) \to 0\ \ \ \ {\text{and}}\ \ \ \ \delta(r\to\infty)
\to 0\
\end{equation}
functional behaviors.

The spatial behavior of these radially-dependent static metric
functions is determined by the Einstein equations
$G^{\mu}_{\nu}=8\pi T^{\mu}_{\nu}$. In particular, denoting the
components of the spherically symmetric energy-momentum tensor by
\cite{Bond1}
\begin{equation}\label{Eq5}
\rho\equiv -T^{t}_{t}\ \ \ \ \text{and}\ \ \ \ p\equiv
T^{r}_{r}=T^{\theta}_{\theta}=T^{\phi}_{\phi}\  ,
\end{equation}
where $\rho$ and $p$ are respectively the energy density and the
isotropic pressure of the spatially regular matter configurations,
one can express the Einstein-matter field equations in the form
\cite{Hodt1,Noteprm}
\begin{equation}\label{Eq6}
C'=8\pi r\rho-{{C}\over{r}}\
\end{equation}
and
\begin{equation}\label{Eq7}
\delta'=-{{4\pi r(\rho +p)}\over{1-C}}\  .
\end{equation}
The components of the energy-momentum tensor which characterize the
static matter configurations are assumed to satisfy the dominant
energy condition \cite{HawEl}
\begin{equation}\label{Eq8}
0\leq |p|\leq\rho\  .
\end{equation}

The gravitational mass contained within a sphere of radius $r$ is
given by the integral relation \cite{Hodt1}
\begin{equation}\label{Eq9}
m(r)=4\pi\int_{0}^{r} x^{2} \rho(x)dx\  ,
\end{equation}
which, taking cognizance of Eq. (\ref{Eq6}), yields the relation
\cite{Hodt1}
\begin{equation}\label{Eq10}
C(r)={{2m(r)}\over{r}}\
\end{equation}
for the characteristic radially dependent dimensionless compactness
parameter of the spatially regular self-gravitating matter
configurations.

Below we shall analyze the spatial behavior of the dimensionless
isotropic pressure function
\begin{equation}\label{Eq11}
P(r)\equiv r^2p(r)\  .
\end{equation}
Taking cognizance of Eqs. (\ref{Eq8}), (\ref{Eq9}), and
(\ref{Eq11}), one finds that, for spatially regular asymptotically
flat spacetimes, this radially dependent pressure function is
characterized by the simple small-$r$ \cite{Hodt1}
\begin{equation}\label{Eq12}
P(r\to0)\to0\
\end{equation}
and large-$r$ \cite{Hodt1}
\begin{equation}\label{Eq13}
rP(r\to\infty)\to0\
\end{equation}
functional behaviors. Substituting the Einstein differential
equations (\ref{Eq6}) and (\ref{Eq7}) into the conservation equation
\begin{equation}\label{Eq14}
T^{\mu}_{r ;\mu}=0\  ,
\end{equation}
one obtains the gradient relation \cite{Hodt1}
\begin{equation}\label{Eq15}
{{2}\over{r}}P'(r)={{{\cal R}(\rho+p)}\over{1-C}}+2(-\rho+p)\
\end{equation}
for the dimensionless pressure function (\ref{Eq11}), where
\begin{equation}\label{Eq16}
{\cal R}(r)\equiv 2-3C-8\pi P\  .
\end{equation}

\section{Null circular geodesics of spherically symmetric curved spacetimes}

The characteristic radial equation for the null circular geodesics
of spherically symmetric curved spacetimes has been derived in
\cite{Chan,CarC,Hodt1}. For completeness of the presentation, we
shall give in the present section a brief sketch of the derivation.
We first note that, as explicitly proved in
\cite{Chan,CarC,Hodt1}, the null circular geodesics of the
spherically symmetric spacetime (\ref{Eq2}) are characterized by the
relations \cite{Notethr}
\begin{equation}\label{Eq17}
V_r=E^2\ \ \ \ \text{and}\ \ \ \ V'_r=0\  ,
\end{equation}
where the effective radial potential that governs the null
trajectories is given by \cite{Chan,CarC,Hodt1}
\begin{equation}\label{Eq18}
V_r=(1-e^{2\delta})E^2+(1-C){{L^2}\over{r^2}}\ .
\end{equation}
Here the conserved physical quantities $\{E,L\}$ are respectively
the energy and the angular momentum along the null trajectories
\cite{Chan,CarC,Hodt1}.

Taking cognizance of the Einstein equations (\ref{Eq6}) and
(\ref{Eq7}), one obtains from Eqs. (\ref{Eq17}) and (\ref{Eq18}) the
radial equation \cite{Chan,CarC,Hodt1}
\begin{equation}\label{Eq19}
{\cal R}(r=r_{\gamma})=0\
\end{equation}
which, for compact enough matter configurations [see Eq.
(\ref{Eq28}) below], determines the discrete radii of the null
circular geodesics (light rings) that characterize the spherically
symmetric curved spacetime (\ref{Eq2}). Interestingly, it has
recently been proved \cite{Hodt0,CBH,Hodmz2} that spatially regular
horizonless spacetimes generally possess an even number of light
rings \cite{Noteift2}.

\section{Lower bound on the local compactness parameter at the outer light ring}

In the present section we shall explicitly prove that, using
analytical techniques, one can derive a lower bound on the
dimensionless compactness parameter $C(r^{\text{out}}_{\gamma})$
[see Eq. (\ref{Eq10})] which characterizes the outermost null
circular geodesic (outermost light ring) of the spatially regular
isotropic ultra-compact objects.

We first point out that, taking cognizance of Eqs. (\ref{Eq3}),
(\ref{Eq4}), (\ref{Eq12}), and (\ref{Eq13}), one finds that the
dimensionless radial function ${\cal R}(r)$ [see Eq. (\ref{Eq16})]
is characterized by the relations
\begin{equation}\label{Eq20}
{\cal R}(r=0)=2\ \ \ \ \text{and}\ \ \ \ {\cal R}(r\to\infty)\to 2\
.
\end{equation}
From Eqs. (\ref{Eq19}) and (\ref{Eq20}) one deduces that the
outermost null circular geodesic of the spherically symmetric horizonless ultra-compact
objects is characterized by the gradient relation
\begin{equation}\label{Eq21}
{\cal R}'(r=r^{\text{out}}_{\gamma})\geq 0\  .
\end{equation}
In addition, taking cognizance of Eqs. (\ref{Eq6}), (\ref{Eq15}),
(\ref{Eq16}), and (\ref{Eq19}), one finds the functional relation
\begin{equation}\label{Eq22}
{\cal R}'(r=r_{\gamma})={{2}\over {r_{\gamma}}}\big[1-8\pi
r^2_{\gamma}(\rho+p)\big]\  ,
\end{equation}
which yields the characteristic inequality [see Eq. (\ref{Eq21})]
\begin{equation}\label{Eq23}
8\pi (r^{\text{out}}_{\gamma})^2(\rho+p)\leq1\
\end{equation}
at the outermost light ring of the ultra-compact matter
configurations.

Taking cognizance of Eqs. (\ref{Eq8}), (\ref{Eq11}), (\ref{Eq16}),
and (\ref{Eq19}), one obtains the dimensionless relations
\cite{Noteprh1}
\begin{equation}\label{Eq24}
C(r^{\text{out}}_{\gamma})={1\over3}\big[2-8\pi(r^{\text{out}}_{\gamma})^2p\big]\geq
{1\over3}\big[2-4\pi(r^{\text{out}}_{\gamma})^2(\rho+p)\big]\  .
\end{equation}
Substituting into (\ref{Eq24}) the characteristic inequality
(\ref{Eq23}), one finds the lower bound
\begin{equation}\label{Eq25}
C(r^{\text{out}}_{\gamma})\geq{{1}\over{2}}\
\end{equation}
on the dimensionless compactness parameter at the outermost light
ring of the spatially regular horizonless ultra-compact objects.

\section{Lower bound on the global compactness parameter of the ultra-compact objects}

In the present section we shall derive a lower bound on the global
compactness parameter $\text{max}_r\{C(r)\}$ which characterizes the
spatially regular isotropic ultra-compact objects. As noted above
[see Eq. (\ref{Eq1})], it has previously been proved that regular
self-gravitating matter configurations with a {\it non}-negative
energy-momentum trace are characterized by the global lower bound
\cite{Hodt0,Noteni}
\begin{equation}\label{Eq26}
\text{max}_r\big\{C(r)\big\}> {2\over3}\ \ \ \ \text{for}\ \ \ \
T\geq0\  .
\end{equation}

We shall now prove that a slightly weaker bound can be derived on
the dimensionless compactness parameter $C(r^{\text{out}}_{\gamma})$
of ultra-compact objects which are characterized by the opposite
({\it negative}) isotropic trace relation \cite{Bond1}
\begin{equation}\label{Eq27}
T=-\rho+3p<0\  .
\end{equation}
Taking cognizance of Eqs. (\ref{Eq11}), (\ref{Eq16}), (\ref{Eq19}),
and (\ref{Eq27}), one obtains the series of dimensionless
inequalities \cite{Noteprh2}
\begin{equation}\label{Eq28}
C(r^{\text{out}}_{\gamma})={1\over3}\big[2-8\pi(r^{\text{out}}_{\gamma})^2p\big]>
{1\over3}\big[2-2\pi(r^{\text{out}}_{\gamma})^2(\rho+p)\big]\geq
{1\over3}\big(2-{1\over4}\big)={{7}\over{12}}\ \ \ \ \ \text{for}\ \
\ \ \ T<0\
\end{equation}
which characterize the outermost null circular geodesic of the
self-gravitating isotropic ultra-compact objects.

\section{Summary}

Spatially regular ultra-compact objects are described by horizonless
curved spacetimes which, like black-hole spacetimes, possess closed
light rings (null circular geodesics). These non-singular compact
objects may provide exotic alternatives to canonical black-hole
spacetimes and their physical properties have therefore been
explored extensively in recent years (see
\cite{Carm,Hodt0,Kei,Mag,Hodmz1,CBH,Hodmz2,Zde1} and references
therein). It should be realized, however, that the characteristic
dimensionless compactness parameter of these exotic matter configurations is
lower than the corresponding compactness parameter $2m(r_{\text{H}})/r_{\text{H}}=1$ of spherically symmetric classical
black-hole spacetimes.

In the present paper we have studied the physical and mathematical
properties of the spatially regular ultra-compact matter
configurations. In particular, we have examined the possibility of
providing  a {\it quantitative} meaning to the important physical
term `ultra-compact object'. We have therefore raised the physically
interesting question: How compact is an ultra-compact object?

Using analytical techniques, we have demonstrated that one can
obtain a non-trivial lower bound on the characteristic compactness
parameter of spherically symmetric ultra-compact objects. In
particular, it has been explicitly proved that horizonless isotropic
ultra-compact matter configurations which possess light rings are
characterized by the compact dimensionless lower bound [see Eqs.
(\ref{Eq10}), (\ref{Eq26}), and (\ref{Eq28})]
\begin{equation}\label{Eq29}
\text{max}_r\Big\{{{2m(r)}\over{r}}\Big\}>{7\over12}
\end{equation}
on their global compactness parameter.

Null circular geodesics (closed light rings) are usually associated
with black-hole spacetimes. Interestingly, though, horizonless
compact objects whose global compactness parameter is characterized
by the sub-black hole relation $\text{max}_r\{2m(r)/r\}<1$
\cite{Notecbh} may also possess null circular geodesics
\cite{Carm,Hodt0,Kei,Mag,Hodmz1,CBH,Hodmz2,Zde1}. In the present
analysis we have explicitly proved that, under the assumptions of
spherical symmetry and isotropy, closed light rings cannot be
associated with arbitrarily dilute matter configurations. In
particular, taking cognizance of the analytically derived lower
bound (\ref{Eq29}) and the recently estimated compactness parameter
of neutron stars \cite{ChPk}, one deduces that astrophysically
realistic isotropic neutron stars cannot possess light rings.

\bigskip
\noindent {\bf ACKNOWLEDGMENTS}

This research is supported by the Carmel Science Foundation. I would
also like to thank Yael Oren, Arbel M. Ongo, Ayelet B. Lata, and
Alona B. Tea for stimulating discussions.


\begin{thebibliography}{99}
\bibitem{Notetr} The trace of a spherically symmetric
energy-momentum tensor can be expressed in the form
$T=-\rho+p_r+2p_{\text{T}}$, where $\{\rho,p_r,p_{\text{T}}\}$ are
respectively the energy density, the radial pressure, and the
tangential pressure of the matter fields \cite{Hodlb,Bond1}.
\bibitem{Noteunits} We shall use natural units in which $G=c=1$.
\bibitem{Notemr} Here $m(r)$ is the gravitational mass of the
horizonless field configuration which is contained within a sphere
of radius $r$ [see Eq. (\ref{Eq9}) below].
\bibitem{Notecbh} Spherically symmetric black-hole spacetimes are
characterized by the well-known compactness relation
$2m(r_{\text{H}})/r_{\text{H}}=1$, where $r_{\text{H}}$ is the
horizon radius of the central black hole.
\bibitem{Noteift1} It is worth noting that these spatially regular
horizonless spacetimes generally possess an even number of light
rings \cite{Hodt0,CBH}. However, as explicitly proved in
\cite{Hodmz2}, there are special ultra-compact horizonless matter
configurations with degenerate null circular geodesics [these
special light rings are characterized by the relations ${\cal
R}(r=r_{\gamma})={\cal R}'(r=r_{\gamma})=0$, see Eq. (\ref{Eq16})
below] which may possess an odd number of null circular geodesics.
\bibitem{Notesc} Here $(t,r,\theta,\phi)$ are the familiar Schwarzschild spacetime coordinates.
\bibitem{Notethr} Note that these two functional relations, which determine the null circular
geodesics of the spherically symmetric spacetime (\ref{Eq2}),
correspond to $\dot r^2=(\dot r^2)'=0$ \cite{Chan,CarC,Hodt1}, where
a dot $^{.}$ denotes a derivative with respect to an affine
parameter.
\bibitem{Noteift2} As emphasized above, it has been explicitly demonstrated in
\cite{Hodmz2} that there are special ultra-compact objects with
degenerate null circular geodesics [which are characterized by the
degenerate relation ${\cal R}(r=r_{\gamma})={\cal
R}'(r=r_{\gamma})=0$] that may possess an odd number of light rings.
\bibitem{Noteprh1} Here we have used the characteristic
inequality $p\leq\rho$ [see (\ref{Eq8})], which yields the relation
$p\leq(\rho+p)/2$.
\bibitem{Noteni} It is worth noting that the global lower bound (\ref{Eq26})
has been derived in \cite{Hodt0} without using the isotropy assumption.
\bibitem{Noteprh2} Here we have used the characteristic
inequality $p<\rho/3$, which yields the relation $p<(\rho+p)/4$ for
isotropic matter configurations that respect the negative trace
relation (\ref{Eq27}).
\end{thebibliography}
\end{document}